\newcommand{\gray}{\mbox{$\gamma$-ray}}

\newcommand{\pcmq}{\mbox{cm$^{-2}$}}
\newcommand{\psec}{\mbox{s$^{-1}$}}
\newcommand{\pkev}{\mbox{keV$^{-1}$}}
\newcommand{\funit}{\mbox{ph \pcmq \psec}}
\newcommand{\feunit}{\mbox{ph \pcmq \psec \pkev}}
\def\la{\mathrel{\mathchoice {\vcenter{\offinterlineskip\halign{\hfil
$\displaystyle##$\hfil\cr<\cr\sim\cr}}}
{\vcenter{\offinterlineskip\halign{\hfil$\textstyle##$\hfil\cr
<\cr\sim\cr}}}
{\vcenter{\offinterlineskip\halign{\hfil$\scriptstyle##$\hfil\cr
<\cr\sim\cr}}}
{\vcenter{\offinterlineskip\halign{\hfil$\scriptscriptstyle##$\hfil\cr
<\cr\sim\cr}}}}}
\def\ga{\mathrel{\mathchoice {\vcenter{\offinterlineskip\halign{\hfil
$\displaystyle##$\hfil\cr>\cr\sim\cr}}}
{\vcenter{\offinterlineskip\halign{\hfil$\textstyle##$\hfil\cr
>\cr\sim\cr}}}
{\vcenter{\offinterlineskip\halign{\hfil$\scriptstyle##$\hfil\cr
>\cr\sim\cr}}}
{\vcenter{\offinterlineskip\halign{\hfil$\scriptscriptstyle##$\hfil\cr
>\cr\sim\cr}}}}}

\def\deg{\ensuremath{^\circ}}

\documentclass[a4paper]{spie}
\usepackage{graphicx}

\title{GRI: the Gamma-Ray Imager mission}

\author{J\"urgen Kn\"odlseder\supit{a} (on behalf of the GRI 
        consortium\supit{b})
\skiplinehalf
\supit{a}Centre d'Etude Spatiale des Rayonnements, 
	 9, avenue du Colonel-Roche, 
	 31028 Toulouse,
	 France;\\
\supit{b}
APC (France),
Argonne National Laboratory (United States),
CESR (France),
CNM (Spain),
Copernicus Astronomical Center (Poland),
CSNSM (France),
DNSC (Denmark),
IAP (France),
IEEC/CSIC (Spain),
INAF Brera (Italy),
IFAE (Spain),
INAF-IASF Bologna (Italy),
INAF-IASF Milano (Italy),
INAF-IASF Palermo (Italy),
INAF-IASF Roma (Italy),
INAF-OAR (Italy),
IOFFE (Russia),
ILL (France),
LAM (France),
MPE (Germany),
Mullard Space Science Laboratory (United Kindom),
SINP, MSU (Russia),
SRON (The Netherlands),
SSL Berkeley (United States),
Univ.~of Coimbra (Portugal),
Univ.~of Ferrara (Italy),
Univ.~of Southampton (United Kingdom),
Univ.~of Utrecht (The Netherlands)
}

\authorinfo{Further author information: (Send correspondence to 
J.K.)\\J.K.: E-mail: knodlseder@cesr.fr, Telephone: +33 (0)5 61 55 66 63}

\begin{document}
\maketitle

\begin{abstract}

Observations of the gamma-ray sky reveal the most powerful sources and 
the most violent events in the Universe. 
While at lower wavebands the observed emission is generally dominated 
by thermal processes, the gamma-ray sky provides us with a view on the 
non-thermal Universe. 
Here particles are accelerated to extreme relativistic energies by 
mechanisms which are still poorly understood, and nuclear reactions are 
synthesizing the basic constituents of our world. 
Cosmic accelerators and cosmic explosions are the major science themes 
that are addressed in the gamma-ray regime.

With the INTEGRAL observatory, ESA has provided a unique tool to the 
astronomical community revealing 
hundreds of sources, 
new classes of objects, 
extraordinary views of antimatter annihilation in our Galaxy, 
and fingerprints of recent nucleosynthesis processes.
While INTEGRAL provides the global overview over the soft gamma-ray 
sky, there is a growing need to perform deeper, more focused investigations 
of gamma-ray sources. 
In soft X-rays a comparable step was taken going from the Einstein 
and the EXOSAT satellites to the Chandra and XMM/Newton observatories. 
Technological advances in the past years in the domain of gamma-ray 
focusing using Laue diffraction and multilayer-coated mirror techniques 
have paved the way towards a gamma-ray mission, providing major 
improvements compared to past missions regarding sensitivity and angular 
resolution. 
Such a future Gamma-Ray Imager will allow to study particle acceleration 
processes and explosion physics in unprecedented detail, providing essential
clues on the innermost nature of the most violent and most energetic processes 
in the Universe.
\end{abstract}
    

\section{WHY GAMMA-RAY ASTRONOMY?}

As introductory remark, it is worth emphasizing some unique features of 
gamma-ray astronomy:
the specific character of the emission processes,
the diversity of the emission sites, and
the penetrating nature of the emission.

First, the emission process that leads to gamma-rays is in general very 
specific, and as such, is rarely observable in other wavebands.
At gamma-ray energies, cosmic acceleration processes are dominant, 
while in the other wavebands thermal processes are generally at the 
origin of the emission.
For example, electrons accelerated to relativistic energies radiate 
gamma-ray photons of all energies through electromagnetic interactions 
with nuclei, photons, or intense magnetic fields.
Accelerated protons generate secondary particles through nuclear interactions, 
which may decay by emission of high-energy gamma-ray photons.
At gamma-ray energies, nuclear deexcitations lead to a manifold of 
line features, while in the other wavebands, it is the bound electrons 
that lead to atomic or molecular transition lines.
For example, the radioactive decay of tracer isotopes allows the 
study of nucleosynthesis processes that occur in the deep inner layers of 
stars.
The interaction of high-energy nuclei with the gas of the 
interstellar medium produces a wealth of excitation lines that probe 
the composition and energy spectrum of the interacting particles.
Finally, annihilation between electrons and positrons result in a unique 
signature at 511~keV that allows the study of antimatter in the Universe.

Second, the sites of gamma-ray emission in the Universe are very 
diverse, and reach from the nearby Sun up to the distant Gamma-Ray 
Bursts and the cosmic background radiation.
Cosmic acceleration takes place on all scales:
locally in solar flares, within our Galaxy (e.g.~in compact binaries, 
pulsars and supernova remnants), and also in distant objects (such as 
active galactic nuclei or gamma-ray bursts).
Cosmic explosions are another site of prominent gamma-ray emission.
They produce a wealth of radioactive isotopes, are potential sources 
of antimatter, and accelerate particles to relativistic energies.
Novae, supernovae and hypernovae are thus prime targets of gamma-ray 
astronomy.

Third, gamma-rays are highly penetrating, allowing the study of otherwise 
obscured regions.
Examples are regions of the galactic disk hidden by dense interstellar 
clouds, or the deeper, inner, zones of some celestial bodies, where the 
most fundamental emission processes are at work.
New classes of sources become visible in the gamma-ray domain, that 
are invisible otherwise.

In summary, gamma-ray astronomy provides a unique view of our Universe.
It unveils specific emission processes, a large diversity of emission 
sites, and probes deeply into the otherwise obscured high-energy 
engines of our Universe.
The gamma-ray Universe is the Universe of particle acceleration and 
nuclear physics, of cosmic explosions and non-thermal phenomena.
Exploring the gamma-ray sky means exploring this unique face of our 
world, the face of the evolving violent Universe.

\section{COSMIC ACCELERATORS}
\label{sec:accelerators}

\subsection{The link between accretion and ejection}

As a general rule, accretion in astrophysical systems is often 
accompanied by mass outflows, which in the high-energy domain take the 
form of (highly) relativistic jets.
Accreting objects are therefore powerful particle accelerators, that 
can manifest on the galactic scale as microquasars, or on the 
cosmological scale, as active galactic nuclei, such as Seyfert 
galaxies and Blazars.

Although the phenomenon is relatively widespread, the jet formation 
process is still poorly understood.
It is still unclear how the energy reservoir of an accreting system 
is transformed in an outflow of relativistic particles.
Jets are not always persistent but often transient phenomena, and it 
is still not known what triggers the sporadic outbursts in accreting 
systems.
Also, the collimation of the jets is poorly understood, and in 
general, the composition of the accelerated particle plasma is not 
known (electron-ion plasma, electron-positron pair plasma).
Finally, the radiation processes that occur in jets are not well 
established.

\begin{figure}[!t]
  \begin{center}
    \begin{tabular}{cc}
      \includegraphics[width=9.0cm]{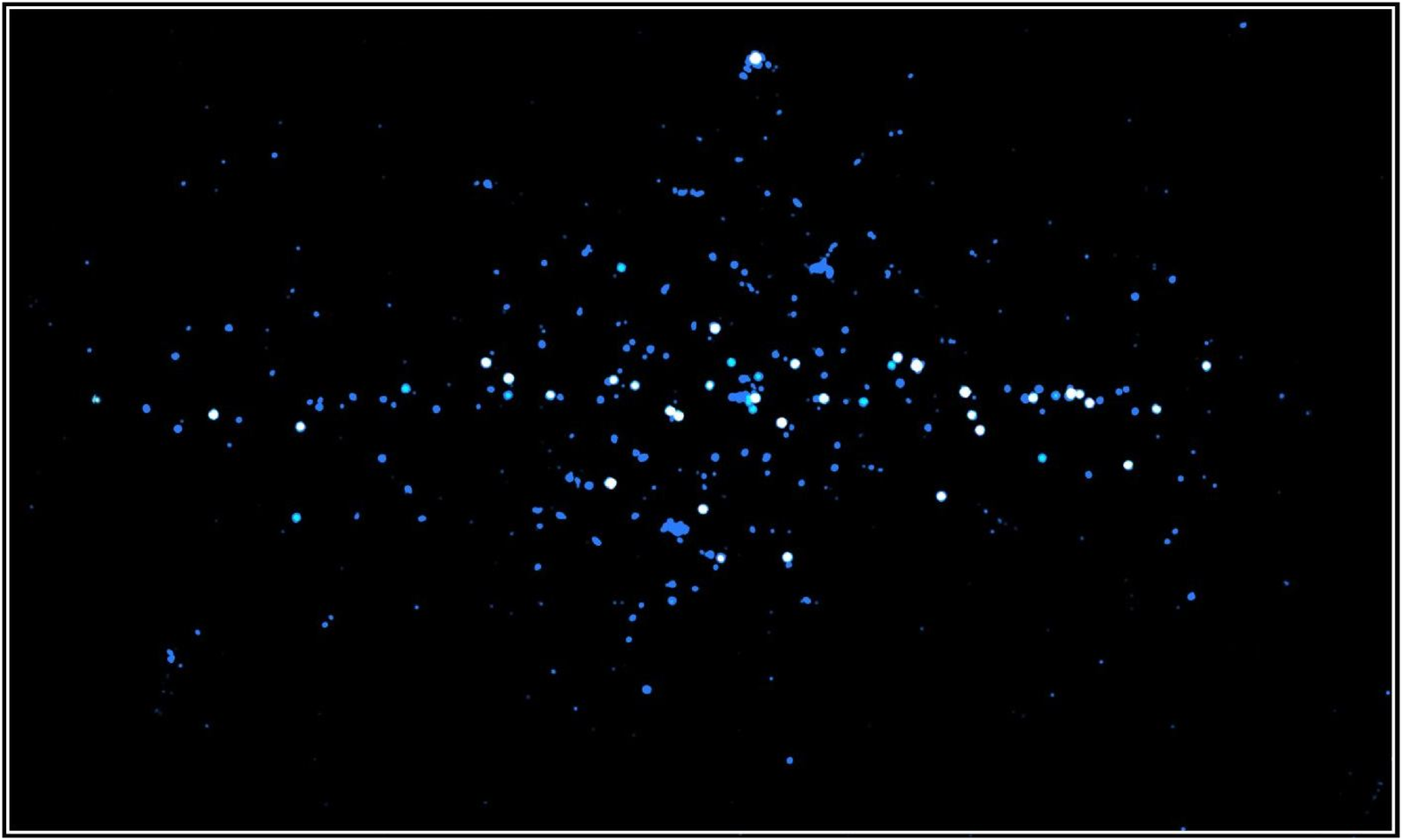}
      \includegraphics[width=7.0cm]{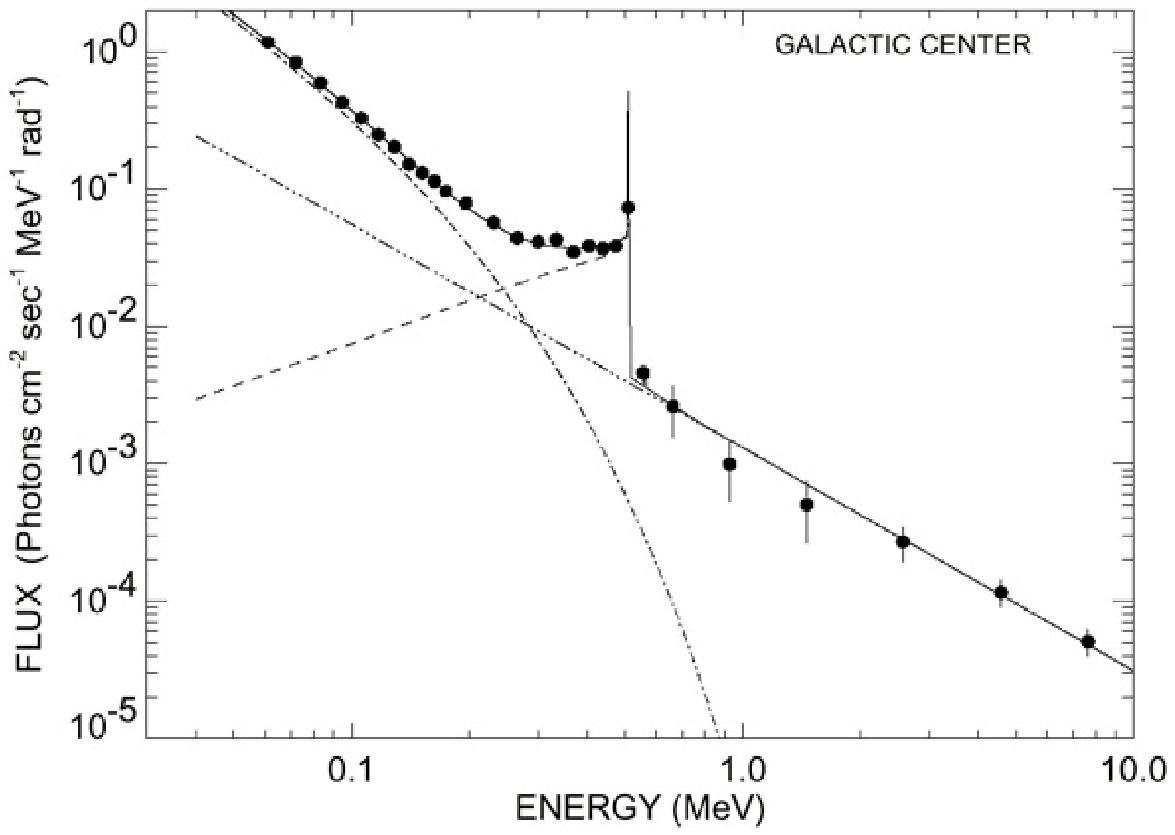}
    \end{tabular}
  \end{center}
  \caption{
    {\em Left panel:} INTEGRAL view of the Galactic Centre region in
    the $20-60$ keV hard X-ray band. 91 sources have been detected
    in this area, resolving $\sim90\%$ of the total flux into 
    individual point sources\cite{lebrun04}.
    {\em Right panel:} OSSE hard X-ray and soft $\gamma-$ray spectrum of
    the Galactic Centre region\cite{kinzer01}. The spectrum is 
    explained by 3 components: an exponentially cut-off powerlaw 
    dominating below $\sim300$ keV, a powerlaw ($\sim E^{-1.7}$) 
    dominating above $511$~keV, and a triangular-shaped positronium 
    continuum component plus a narrow line at $511$~keV.
    \label{fig:gammasky}}
\end{figure}

Observations in the gamma-ray domain are able to provide a number of 
clues to these questions.
Gamma-rays probe the innermost regions of the accreting systems 
that are not accessible in other wavebands, providing the closest 
view to the accelerating engine.
Time variability and polarization studies provide important insights 
into the physical processes and the geometry that govern the 
acceleration site.
The accelerated plasma may reveal its nature through 
characteristic nuclear and/or annihilation line features which may 
help to settle the question about the nature of the accelerated plasma.

\subsection{The origin of galactic soft \gray\ emission}

Since decades, the nature of the galactic hard X-ray ($>15$ keV)
emission has been
one of the most challenging mysteries in the field.
The INTEGRAL imager IBIS has now finally solved this puzzle.
At least $90\%$ of the emission has been resolved into point sources, 
settling the debate about the origin of the emission
\cite{lebrun04} (see left panel of Fig.~\ref{fig:gammasky}).

At higher energies, say above $\sim300$ keV, the situation is less clear.
In this domain, only a small fraction of the galactic emission has so far 
been resolved into point sources, and the nature of the bulk of the 
galactic emission is so far unexplained.
That a new kind of object or emission mechanism should be at work in 
this domain is already suggested by the change of the slope of 
the galactic emission spectrum (see right panel of Fig.~\ref{fig:gammasky}).
While below $\sim300$ keV the spectrum can be explained by a 
superposition of Comptonisation spectra from individual point 
sources, the spectrum turns into a powerlaw above this energy, which 
is reminiscent of particle acceleration processes.
Identifying the source of this particle acceleration process, i.e. 
identifying the origin of the galactic soft gamma-ray emission, is 
one of the major goals of a future gamma-ray mission.

One of the strategies to resolve this puzzle is to follow the 
successful road shown by INTEGRAL for the hard X-ray emission: trying 
to resolve the emission into individual point sources.
Indeed, a number of galactic sources show powerlaw spectra in the 
gamma-ray band, such as supernova remnants, like the Crab nebula, or 
some of the black-hole binary systems, like Cyg X-1
\cite{mcconnell00}.
Searching for the hard powerlaw emission tails in these objects is 
therefore a key objective for a future gamma-ray mission.

\subsection{The origin of the soft \gray\ background}

\begin{figure}[t!]
  \begin{center}
    \begin{tabular}{c}
      \includegraphics[width=16cm]{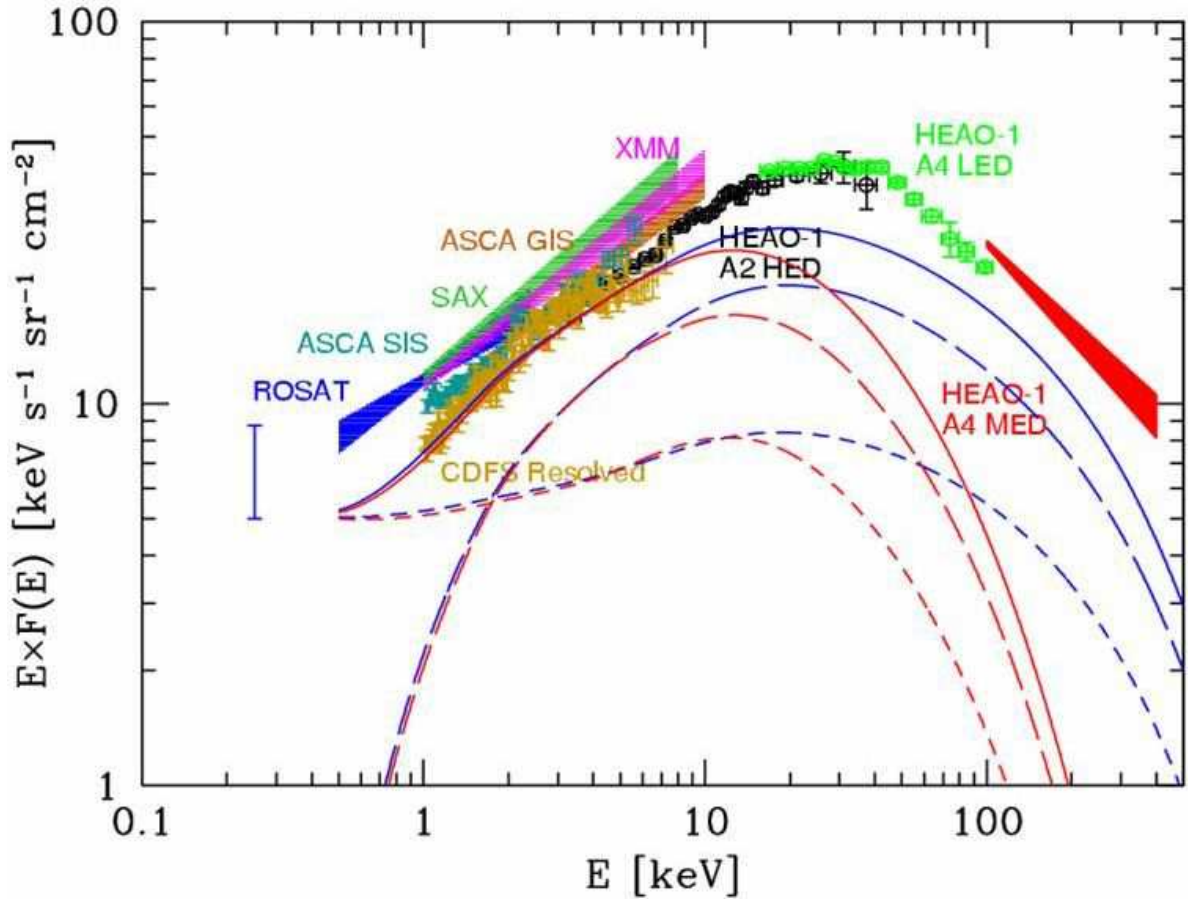}
    \end{tabular}
  \end{center}
  \caption{The $0.25-400$ keV cosmic background spectrum
     fitted with synthesis models\cite{comastri04}. 
     None of the models provides a satisfactory fit of the 
     observations.
    \label{fig:agn}}
\end{figure}

After the achievements of XMM-Newton and Chandra, the origin of the
cosmic X-ray background (CXB) is now basically solved for energies close 
to a few keV.
Below 8~keV about $\ga 80\%$ of the emission has been resolved into 
individual sources, which have been identified as active galactic 
nuclei (AGN)\cite{hasinger04}.
Above $\sim8$~keV, however, only $\la 50\%$ of the CXB has been 
resolved into sources\cite{worsley05}, while in the $20-100$~keV 
hard X-ray band, where IBIS is most sensitive, only $\sim1\%$ of the 
emission has been resolved\cite{bassani06}.
Above this energy, in the soft \gray\ band, basically nothing is know 
about the nature of the cosmic background radiation.

While the situation in the hard X-ray band ($\la 100$~keV) may change 
after the launch of the Simbol-X telescope (see contribution of 
P.~Ferrando, these proceedings), the soft \gray\ band remains unexplored.
It is however this energy band which may provide the key for the 
understanding of the cosmic background radiation.
Synthesis models, which are well established and tested against observational 
results, can be used to evaluate the integrated AGN contribution to the 
soft \gray\ background. 
However, the spectral shape of the different classes of AGN that are 
used for modelling the background has so far not been firmly established 
at soft \gray\ energies. 
As an illustration, Fig.~\ref{fig:agn} shows the impact of the AGN 
power law cut-off energy on the resulting prediction of the cosmic 
background radiation.
Observations by BeppoSAX\cite{risaliti02,perola02} of a 
handful of radio quiet sources, loosely locate this drop-off in the range 
$30-300$ keV; furthermore these measurements give evidence for a variable 
cut-off energy and suggest that it may increase with 
increasing photon index\cite{perola02}. 
In radio loud sources the situation is even more complicated with some objects
showing a power law break and others no cut-off up to the MeV region. 
In a couple of low luminosity AGN no cut-off is present up to 
$300-500$~keV. 
The overall picture suggests some link with the absence 
(low energy cut-off) or presence (high energy cut-off)  
of jets in the various AGN types sampled, but the data are still too scarce 
for a good understanding of the processes involved. 

Therefore, the goal of GRI is to measure the soft \gray\ Spectra Energy 
Distribution (SED) in a sizeable fraction of AGN in order to determine 
average shapes in individual classes and so the nature of the radiation 
processes at the heart of all AGN.
This would provide at the same time information for soft \gray\ 
background synthesis models.
On the other hand, sensitive deep field observations should be able to 
resolve the soft \gray\ background into individual sources, allowing 
for the ultimate identification of the origin of the emission.

\subsection{Particle acceleration in extreme B-fields}

\begin{figure}[!t]
  \begin{center}
    \begin{tabular}{c}
      \includegraphics[width=8cm]{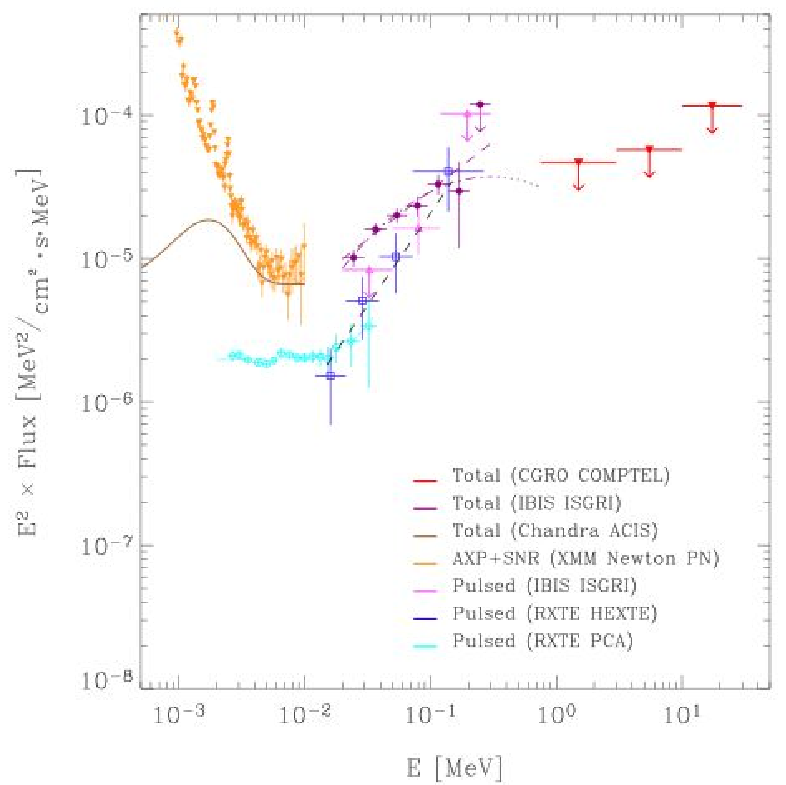}
      \includegraphics[width=8cm]{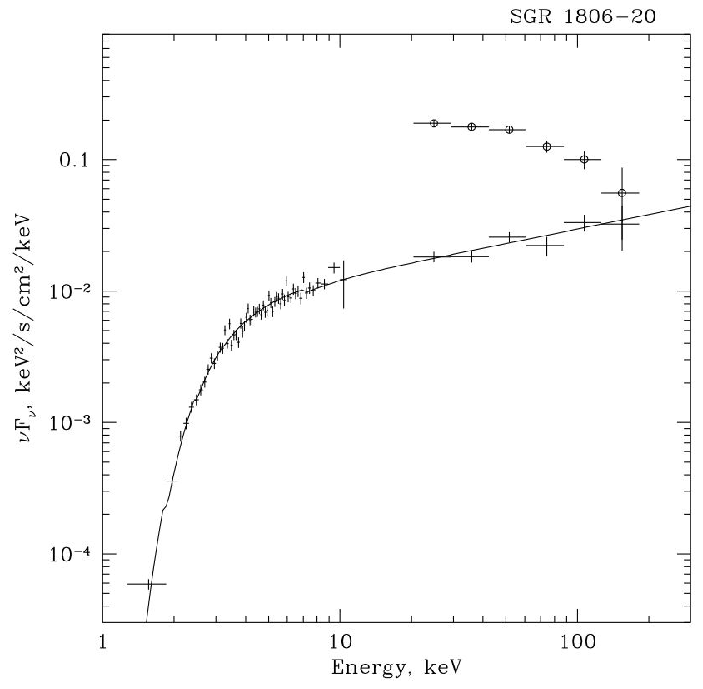}
    \end{tabular}
  \end{center}
  \caption{
    {\em Left panel:} High-energy emission spectrum
    from the AXP 1E~1841-045 and the surrounding SNR Kes~73
    \cite{kuiper06}.
    The COMPTEL upper limits indicate that the spectra should break 
    between $150 - 1000$ keV.
    {\em Right panel:} The quiescent energy spectrum of SGR~1806-20 (lower 
    spectrum) and the summed spectrum of all detected bursts rescaled 
    down by a factor of 1000 (upper spectrum) \cite{molkov05}.
    So far, no indication for a spectral break above $100$ keV are seen.
    \label{fig:sgr}}
\end{figure}

The strong magnetic fields that occur at the surface of neutron stars 
in combination with their fast rotation make them to powerful 
electrodynamic particle accelerators, which may manifest as pulsars 
to the observer.
Gamma-ray emitting pulsars can be divided into 3 classes:
spin-down powered pulsars, such as normal isolated pulsars,
accretion powered pulsars, occurring in low-mass or high-mass binary 
systems, like spin-up/spin-down and millisecond pulsars,
and magnetically powered pulsars, known as magnetars.

Despite the longstanding efforts in understanding the physics of 
spin-down powered pulsars, the site of the gamma-ray production within 
the magnetosphere (outer gap or polar cap) and the physical process at 
action (synchrotron emission, curvature radiation, inverse Compton 
scattering) remain undetermined.
Although most of the pulsars are expected to reach their maximum 
luminosity in the MeV domain, the relatively weak photon fluxes have 
only allowed the study of a handful of objects so far.
Increasing the statistics will enable the study of the pulsar
lightcurves over a much broader energy range than today, providing 
crucial clues on the acceleration physics of these objects.

Before the launch of INTEGRAL, the class of anomalous X-ray pulsars 
(AXPs), suggested to form a sub-class of the magnetar population, were 
believed to exhibit very soft X-ray spectra.
This picture, however, changed dramatically with the detection of 
AXPs in the soft gamma-ray band by INTEGRAL \cite{kuiper04,kuiper06}.
In fact, above $\sim10$ keV a dramatic upturn is observed in the 
spectra which is expected to cumulate in the 100 keV -- 1 MeV domain
(see Fig.~\ref{fig:sgr}).
The same is true for Soft Gamma-ray Repeaters (SGRs), as illustrated 
by the recent discovery of quiescent soft gamma-ray emission from 
SGR~1806-20 by INTEGRAL \cite{molkov05} (c.f.~Fig.~\ref{fig:sgr}).
The process that gives rise to the observed gamma-ray emission in 
still unknown.
No high-energy cut-off has so far been observed in the spectra, yet 
upper limits in the MeV domain indicate that such a cut-off should be 
present.
Determining this cut-off may provide important insights in the 
physical nature of the emission process, and in particular, about the 
role of QED effects, such as photon splitting, in the extreme 
magnetic field that occur in such objects.
Strong polarization is expected for the high-energy emission from 
these exotic objects, and polarization measurements may be crucial 
to disentangle the nature of the emission process and the geometry of 
the emitting region.
Complementary measurements of cyclotron features in the spectra 
provide the most direct measure of the magnetic field strengths, 
complementing our knowledge of the physical parameters of the systems.

\subsection{Particle acceleration in low B-fields}

The possibility to have high  energy emission from cosmic sources
in the range above $0.1-1$ TeV has been argued in the last 20 years.
Recently, HESS (High Energy Stereoscopic System), a ground-based
Cerenkov array telescope has become fully operational and has
performed the first Galactic plane scan with a sensitivity of a
few percent of the Crab at energies above 100 GeV. 
This survey revealed the existence of a population of fourteen TeV 
objects, most of which previously unknown\cite{aharonian05,aharonian06}.
More recently, the MAGIC (Major Atmospheric Gamma Imaging Cerenkov
telescope) collaboration has reported a positive observations of
HESS J1813-178, resulting in a gamma-ray flux consistent with the
previous HESS detection and showing a hard power law with 
$\Gamma = 2.1$ in the range from $0.4-10$ TeV\cite{albert06}. 
These findings have an important astrophysical implication for the 
understanding of cosmic particle accelerators via gamma ray measurements 
in order to disentangle the mechanisms active in the different
emitting regions and, in turn, to understand the nature of the sources. 
In fact, the detection of a substantial number of very high energy
Galactic sources emitting a large fraction of energy in the GeV to
TeV range has opened a new space window for astrophysical studies
related to cosmic particle accelerators. 
Different types of galactic sources are known to be cosmic particle 
accelerators and potential sources of high energy gamma rays: 
isolated pulsars and their pulsar wind nebulae (PWN), 
supernova remnants (SNR), 
star forming regions, 
and binary systems with a collapsed object like a microquasar
or a pulsar.  

\begin{figure}[!t]
  \begin{center}
    \begin{tabular}{c}
      \includegraphics[width=8.0cm, height=7cm]{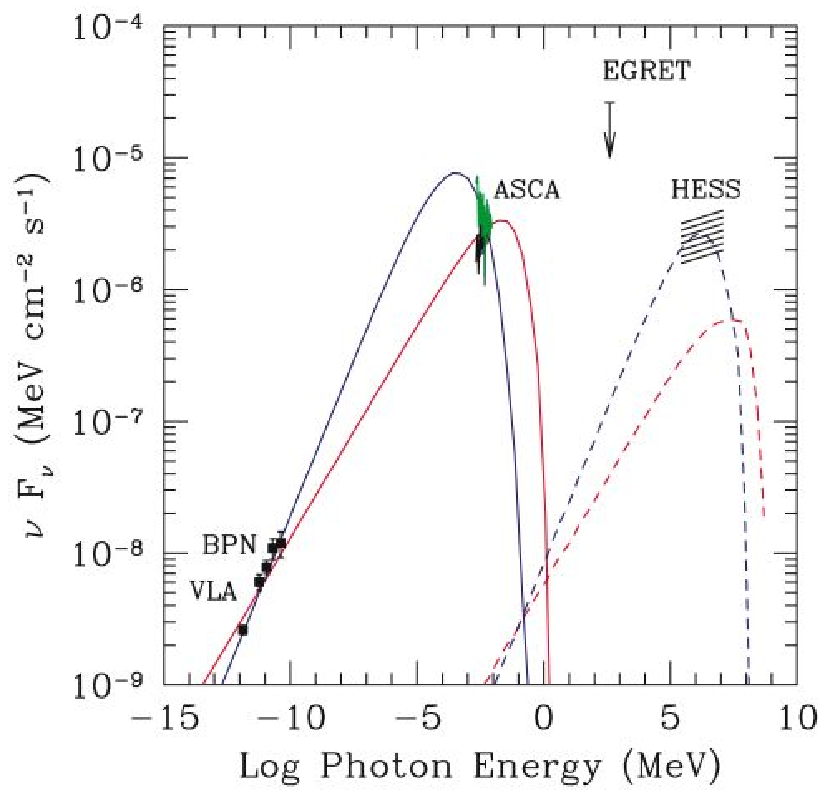}
      \hspace{2.0cm}
      \includegraphics[width=4.5cm, height=7.5cm]{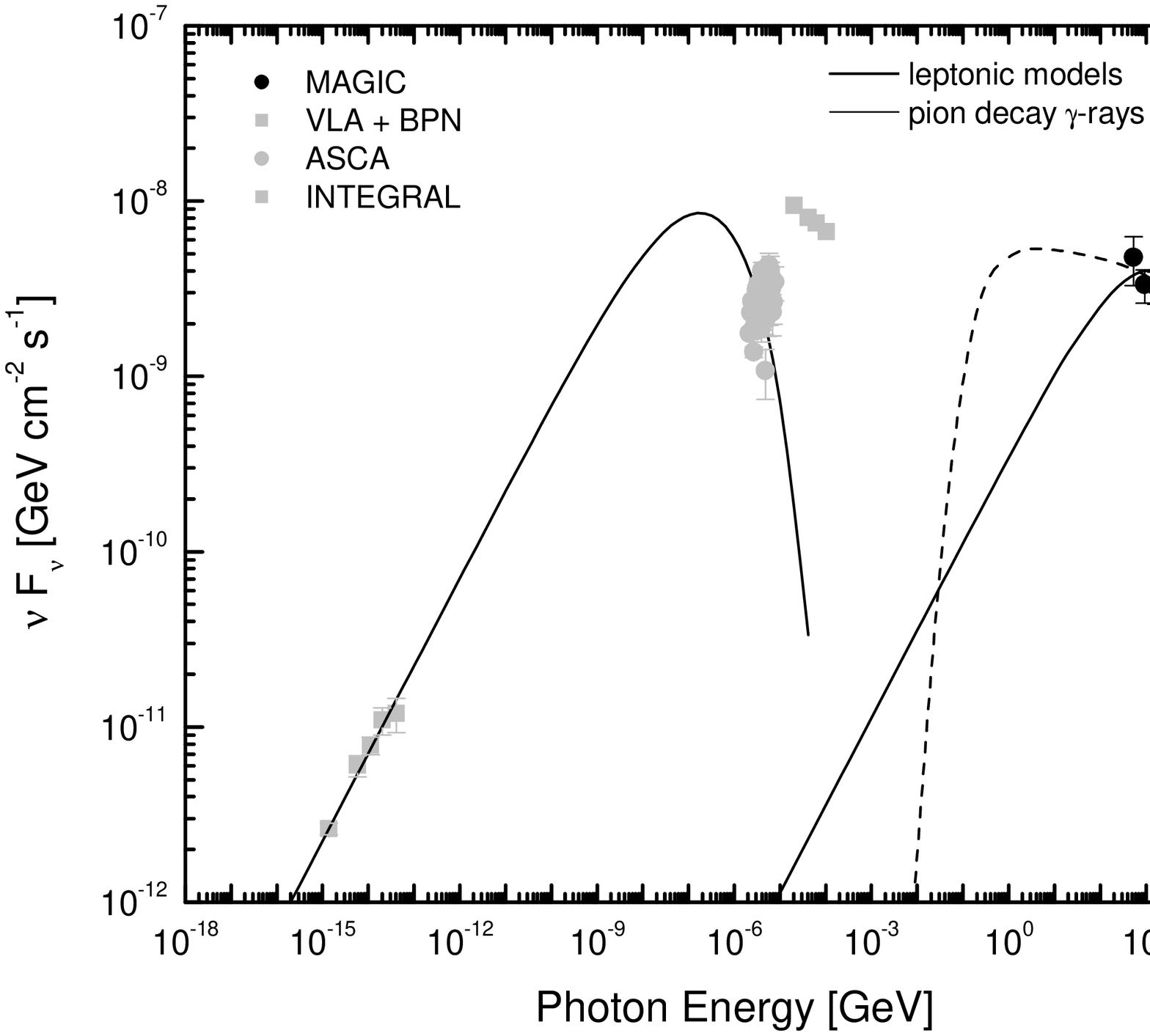}
    \end{tabular}
  \end{center}
  \caption{{\it Left panel}:  Fits to the broadband emission of 
  HESS~J1813-178 assuming that all the flux originates from the 
  shell of SNR G12.8+0.0. 
  The two models indicated by the red and blue lines show the range 
  of parameter space that best fit the data: 
  the red model uses the spectral index from the best fit to the
  ASCA data, $N_{H}$ of $10.8\times10^{22}$ cm$^{-2}$ (black X-ray spectrum), 
  while the blue model uses the spectral index implied by
  the $1\sigma$ lower limit to $N_{H}$ of $8.9\times10^{22}$ cm$^{-2}$ 
  (green X-ray spectrum). 
  Both models include contributions from synchrotron (solid lines) 
  and IC (dashed lines) mechanisms; the filling factor of the magnetic 
  field in the IC emitting region has been assumed to 15\%.
  {\it Right panel}: Leptonic and hadronic models for the HESS~J1813-178 data.
  \label{fig:brogan_albert}}
\end{figure}

So far, out of the HESS sources without any known counterpart only for 
two of them, namely HESS J1813-178 and HESS J1837-067, hard X-ray and 
soft gamma-ray emission has been detected, supporting the Synchrotron 
Inverse Compton scenario\cite{malizia06,ubertini05}. 
Unfortunately, further attempts to model in detail the broad band 
emission from these two objects \cite{brogan05,albert06} have not 
been completely successful\cite{ubertini06}, as shown in 
Fig.~\ref{fig:brogan_albert}. 
In this scenario, the lack of radio/X-ray emission in most of those 
HESS sources is particularly interesting since it strongly suggests 
that the accelerated particles may be nucleons rather than high energy 
electrons.

It is clear that both the hadronic and leptonic models so far proposed
fail to easily explain the whole observational picture if the
radio, hard X-ray, soft gamma-ray and TeV high energy photons are produced
in the same SNR region by a single physical process. 
In fact, to firmly confirm the above hypothesis it is necessary to have
instruments capable to provide arcsec spatially resolved
spectroscopy, not possible with the present generation of 
gamma-ray instruments.

This can be achievable with a new generation of Gamma Ray Imagers
if capable to provide at the same time high angular resolution
coupled with 10 time better sensitivity.

\section{COSMIC EXPLOSIONS}
\label{sec:explosions}

\subsection{Understanding Type~Ia supernovae}

\begin{figure}[!t]
  \begin{center}
    \begin{tabular}{c}
      \includegraphics[width=16cm, height=10cm]{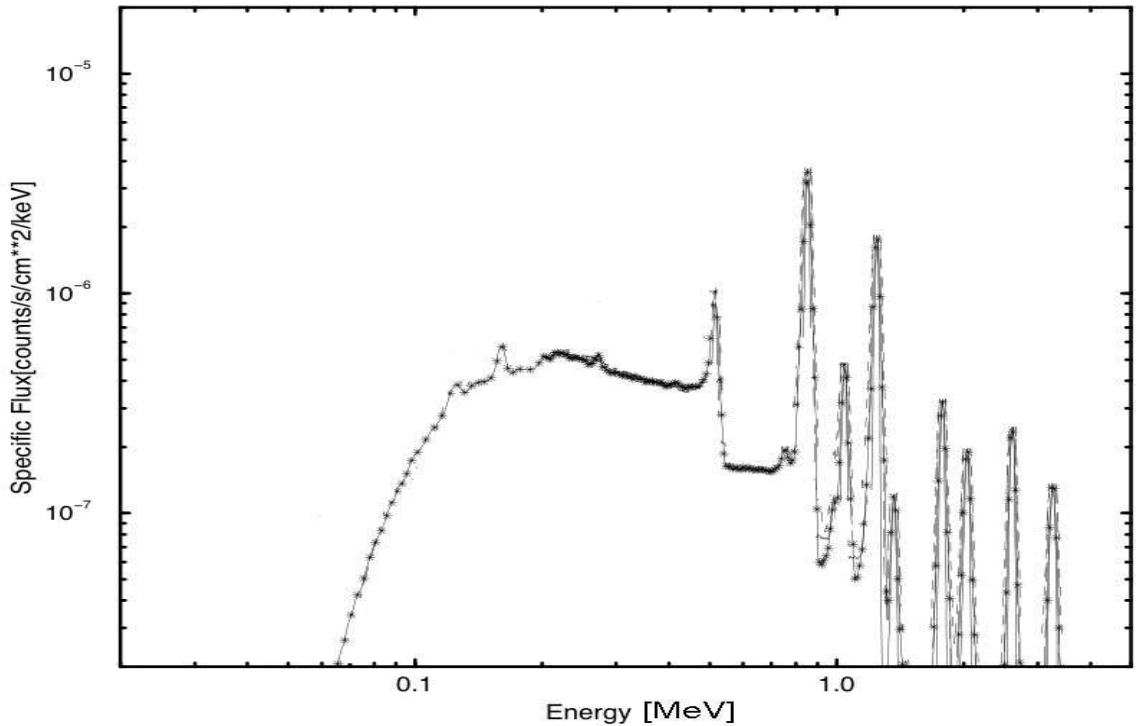}
    \end{tabular}
  \end{center}
  \caption{Simulated gamma-ray spectrum of a Type~Ia supernova
    \cite{gomez98}.
    \label{fig:snia}}
\end{figure}

Although hundreds of Type~Ia supernovae are observed each year, and 
although their optical lightcurves and spectra are studied in great 
detail, the intimate nature of these events is still unknown.
Following common wisdom, Type~Ia supernovae are believed to arise in 
binary systems where matter is accreted from a normal star onto a 
white dwarf.
Once the white dwarf exceeds the Chandrasekhar mass limit a thermonuclear 
runaway occurs that leads to its incineration and disruption.
However, attempts to model the accretion process have so far failed to 
allow for sufficient mass accretion that would push the white dwarf 
over its stability limit \cite{hillebrandt00}.
Even worse, there is no firm clue that Type~Ia progenitors are indeed 
binary systems composed of a white dwarf and a normal star.
Alternatively, the merging of two white dwarfs in a close binary 
system could also explain the observable features of Type~Ia events
\cite{livio03}.
Finally, the explosion mechanism of the white dwarf is only poorly 
understood, principally due to the impossibility to reliably model the 
nuclear flame propagation in such objects \cite{hillebrandt00}.

In view of all these uncertainties it seems more than surprising that 
Type~Ia are widely considered as standard candles.
In particular, it is this standard candle hypothesis that is the 
basis of one of the fundamental discoveries of the last decade: 
the accelerating expansion of the Universe\cite{riess98}.
Although empirical corrections to the observed optical lightcurves 
seem to allow for some kind of standardization, there 
is increasing evidence that Type~Ia supernovae is not an homogeneous 
class of objects \cite{mannucci05}.

Gamma-ray observation of Type~Ia supernovae provide a new and unique 
view of these events.
Nucleosynthetic products of the thermonuclear runaway lead to a rich 
spectrum of gamma-ray line and continuum emission that contains a 
wealth of information on the progenitor system, the explosion 
mechanism, the system configuration, and its evolution
(c.f.~Fig.~\ref{fig:snia}).
In particular, the radioactive decays of $^{56}$Ni and $^{56}$Co, 
which power the optical lightcurve that is so crucial for the 
cosmological interpretation of distant Type~Ia events, can be 
directly observed in the gamma-ray domain, allowing to pinpoint the 
underlying progenitor and explosion scenario.
The comparison of the gamma-ray to the optical lightcurve 
will provide direct information about energy recycling in the 
supernova envelope that will allow a physical (and not only empirical)
calibration of Type~Ia events as standard candles.

In addition to line intensities and lightcurves, the shapes of the 
gamma-ray lines hold important information about the explosion 
dynamics and the matter stratification in the system.
Measuring the line shapes (and their time evolution) will allow to 
distinguish between the different explosion scenarios, ultimately 
revealing the mechanism that creates these most violent events in the 
Universe \cite{gomez98}.

\subsection{Unveiling the origin of galactic positrons}

\begin{figure}[!t]
  \begin{center}
    \begin{tabular}{c}
      \includegraphics[width=16cm]{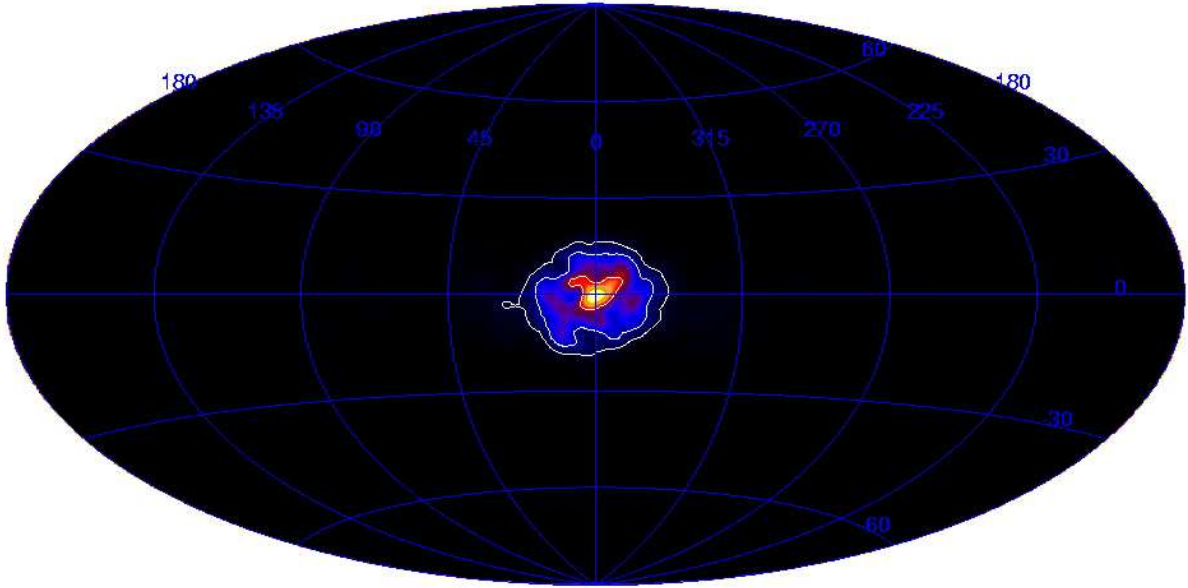}
    \end{tabular}
  \end{center}
  \caption{First all-sky map of 511~keV positron-electron annihilation 
    radiation as observed by the SPI telescope aboard INTEGRAL
    (Kn\"odlseder et al.~2005).
    \label{fig:511keV}}
\end{figure}

The unprecedented imaging and spectroscopy capabilities of the 
spectrometer SPI aboard INTEGRAL have now provided for the first time 
an image of the distribution of 511~keV positron-electron annihilation 
all over the sky \cite{knoedl05} (c.f.~Fig.~\ref{fig:511keV}).
The outcome of this survey is astonishing: 511~keV line emission is 
only seen towards the bulge region of our Galaxy, while the rest of the 
sky remains surprisingly dark.
Only a weak glimmer of 511~keV emission is perceptible from the disk of 
the Galaxy, much less than expected from stellar populations following 
the global mass distribution of the Galaxy.
In other words, positron annihilation seems to be greatly enhanced in 
the bulge with respect to the disk of the Galaxy.

A detailed analysis of the 511~keV line shape measured by SPI has 
also provided interesting insights into the annihilation physics
\cite{churazov05}.
At least two components have been identified, indicating that positron 
annihilation takes place in a partially ionized medium.
This clearly demonstrated that precise 511~keV line shape 
measurements provide important insights into the distribution of the 
various phases of the interstellar medium (ISM)\cite{jean06}.

While INTEGRAL has set the global picture of galactic positron 
annihilation, high angular resolution mapping of the galactic bulge region 
is required to shed light on the still mysterious source of positrons.
So far, no individual source of positron emission could have been 
identified, primarily due to the expected low levels of 511~keV line 
fluxes.
An instrument with sufficiently good sensitivity and angular 
resolution should be able to pinpoint the origin of the positrons, 
by providing detailed maps of the central bulge region of the Galaxy.
With additional fine spectroscopic capabilities, comparable to that 
achieved by the germanium detectors onboard the SPI telescope, the spatial 
variations of the 511~keV line shape will allow to draw an unprecedented 
picture of the distribution of the various ISM phases in the inner regions 
of our Galaxy.

Thus, with the next generation gamma-ray telescope, galactic positrons 
will be exploited as a messenger from the mysterious antimatter source 
in the Milky-Way, as well as a tracer to probe the conditions of the 
ISM that are difficultly to measure by other means.

\subsection{Understanding core-collapse explosions}

Gamma-ray line and continuum observations address some of the most 
fundamental questions of core-collapse supernovae: 
how and where the large neutrino fluxes couple to the stellar ejecta; 
how asymmetric the explosions are, including whether jets form; 
and what are quantitative nucleosynthesis yields from both static and 
explosive burning processes?

The ejected mass of $^{44}$Ti, which is produced in the innermost ejecta 
and fallback matter that experiences the alpha-rich freezeout of nuclear 
statistical equilibrium, can be measured to a precision of several 
percent in SN~1987A. 
Along with other isotopic yields already known, this will provide an 
unprecedented constraint on models of that event. 
$^{44}$Ti can also be measured and mapped, in angle and radial velocity, 
in several historical galactic supernova remnants. 
These measurements will help clarify the ejection dynamics, including 
how common jets initiated by the core collapse are.

Wide-field gamma-ray instruments have shown the global diffuse emission 
from long-lived isotopes $^{26}$Al and $^{60}$Fe, illustrating clearly 
ongoing galactic nucleosynthesis. 
A necessary complement to these are high-sensitivity measurements of the 
yields of these isotopes from individual supernovae. 
A future gamma-ray mission should determine these yields, and map 
the line emission across several nearby supernova remnants, shedding further 
light on the ejection dynamics. 
It is also likely that the nucleosynthesis of these isotopes in hydrostatic 
burning phases will be revealed by observations of individual nearby massive 
stars with high mass-loss rates.

For rare nearby supernovae, within a few Mpc, we will be given a glimpse 
of nucleosynthesis and dynamics from short-lived isotopes $^{56}$Ni, and 
$^{57}$Ni, as was the case for SN~1987A in the LMC. 
In that event we saw that a few percent of the core radioactivity was 
somehow transported to low-optical depth regions, perhaps surprising 
mostly receding from us, but there could be quite some variety, especially 
if jets or other extensive mixing mechanisms are ubiquitous.

\subsection{Nova nucleosynthesis}

Classical novae are another site of explosive nucleosynthesis that is 
still only partially understood\cite{hernanz05}.
Although observed elemental abundances in novae ejecta are relatively 
well matched by theoretical models, the observed amount of matter 
that is ejected substantially exceeds expectations.
How well do we really understand the physics of classical novae?

Radioactive isotopes that are produced during the nova explosion can 
serve as tracer elements to study these\break events.
Gamma-ray lines are expected from relatively long living isotopes, 
such as $^{7}$Be and $^{22}$Na, and from positron annihilation of 
$\beta^{+}$-decay positrons arising from the short living $^{13}$N and 
$^{18}$F isotopes.
Observation of the gamma-ray lines that arise from these isotopes
may improve our insight into the physical processes that govern the 
explosion.
In particular, they provide information on the composition of the white 
dwarf outer layers, the mixing of the envelop during the explosion, and 
the nucleosynthetic yields.
Observing a sizeable sample of galactic nova events in gamma-rays should 
considerably improve our understanding of the processes at work, and 
help to better understand the underlying physics.

\section{THE GRI MISSION}

\subsection{Mission requirements}

The major mission requirement for the future European gamma-ray mission 
is sensitivity.
Many interesting scientific questions are in a domain where photons 
are rare (say $10^{-8}$ \feunit), and therefore large collecting areas are 
needed to perform measurements in a reasonable amount of time.
It is clear that a significant sensitivity leap is required, say 
50--100 times more sensitive than current instruments, if the above 
listed scientific questions should be addressed.

With such a sensitivity leap, the expected number of observable sources 
would be large, implying the need for good angular resolution to avoid 
source confusion in crowded regions, such as for example the galactic 
centre.
Also, it is desirable to have an angular resolution comparable to that 
at other wavebands, to allow for source identification and hence 
multi-wavelength studies.

As mentioned previously, gamma-ray emission may be substantially 
polarized due to the non-thermal nature of the underlying emission 
processes.
Studying not only the intensity and the spectrum but also the polarization
of the emission would add a new powerful scientific dimension to the 
observations.
Such measurements would allow to discriminate between the different 
plausible emission processes at work, and would constrain the geometry 
of the emission sites.

Taking all these considerations into account, the following mission 
requirements derive (c.f.~Table \ref{tab:mission}).
The energy band should cover soft gamma-rays, with an extension down 
to the hard X-ray band to allow the study of broad-band spectra.
At the same time, major gamma-ray lines of astrophysical interest
(i.e.~511~keV and 847~keV) should be accessible.
We therefore set our baseline energy range to $50-2000$ keV.

A real sensitivity leap should be achieved, typical by a factor of 
50--100 with respect to existing gamma-ray instrumentation.
For high-resolution gamma-ray line spectroscopy a good energy resolution 
is desirable to exploit the full potential of line profile studies.
A reasonably sized field-of-view together with arcmin angular 
resolution should allow the imaging of field populations of gamma-ray 
sources in a single observation.
Finally, good polarization capabilities, at the percent level for 
strong sources, are required to exploit this additional observable.

\begin{table}[bht]
  \caption{
    Mission requirements for the future European gamma-ray mission
    (sensitivities are for $10^6$ seconds at $3\sigma$ detection 
    significance).}
  \label{tab:mission}
  \begin{center}
    \leavevmode
    \footnotesize
    \begin{tabular}[h]{ll}
      \hline \\[-5pt]
      Parameter & Requirement \\[+5pt]
      \hline \\[-5pt]
      Energy band             & 50 keV -- 2 MeV \\
      Continuum sensitivity   & $10^{-8}$ \feunit\ \\
      Narrow line sensitivity & $5 \times 10^{-7}$ \funit\ \\
      Energy resolution       & 2 keV at 600 keV \\
      Field of view           & 15 arcmin \\
      Angular resolution      & arcmin \\
      Polarization            & 1\% at 10 mCrab \\
      \hline \\
      \end{tabular}
  \end{center}
\end{table}

How can these mission requirements be reached?
We think that the best solution is the implementation of a broad-band
gamma-ray lens telescope based on the principle of Laue diffraction of 
gamma-rays in mosaic crystals \cite{ballmoos04,halloin04,chiara00}.
The Laue lens may eventually be complemented by a coded mask 
telescope or a multilayer-coated mirror telescope in order to achieve 
coverage in the hard X-ray domain.
In addition, the lens detector may be designed in a Compton 
configuration so that it could be simultaneously used as an all-sky 
monitor.

\subsection{GRI design}

The key element of GRI is a broad-band gamma-ray lens based on the principle 
of Laue diffraction of photons in mosaic crystals (von Ballmoos and 
Frontera et al., these proceedings).
Each crystal can be considered as a little mirror which deviates
\gray s through Bragg reflection from the incident beam onto a focal
spot.
Although the Bragg relation 
\begin{equation}
 2 d \sin \theta = n \frac{h c}{E}
 \label{eq:bragg}
\end{equation}
implies that only a single energy $E$ (and its multiples) can be 
diffracted by a given crystal, the mosaic spread
$\Delta \theta$ that occurs in each crystal leads to an energy spread
$\Delta E \propto \Delta \theta E^2$
($d$ is the crystal lattice spacing, 
$\theta$ the Bragg angle,
$n$ the diffraction order, 
$h$ the Planck constant, 
$c$ the speed of light and 
$E$ the energy of the incident photon).
Placing the crystals on concentric rings around an optical axis, and
the careful selection of the inclination angle on each of the rings,
allows then to build a broad-band gamma-ray lens that has continuous 
energy coverage over a specified band.
Since larger energies $E$ imply smaller diffraction angles $\theta$
(see Eq.~\ref{eq:bragg}), crystals diffracting large energies are 
located on the inner rings of the lens.
Conversely, smaller energies $E$ imply larger diffraction angles and 
consequently are located on the outer rings.

A Laue lens starts to become efficient from energies on where photon
absorption becomes small within the crystals.
For copper crystals and a minimum machinable crystal thickness of 
$1-2$~mm this energy is situated around $E_{\min}\sim100$~keV.
To cover a continuous energy band, the crystal energy bandpass $\Delta E$
should not be too small.
Since $\Delta E \propto E^2$, the number of required crystal tiles
increases considerably with decreasing energy, leading to a natural 
lower energy boundary (also situated around $\sim100$~keV) where Bragg 
reflection becomes inappropriate for photon concentration.

Having these considerations in mind, we propose a broad-band Laue lens 
covering the energy band $\approx 150$~keV -- 2~MeV as the central element 
of GRI.
The lens is composed of a ring with inner and outer diameters of
$\sim1$~m and $\sim3.8$~m, respectively, resulting in
$\sim10$~m$^2$ of geometrical lens area.
Eventually, deployable lens petals may be added to reach even lower 
energies, and we actually are studying the technical feasibility of 
such a solution (Frontera et al., these proceedings).
Owing to the small scatter angles of $\sim0.1\deg - 1\deg$ (for the
largest and smallest energies, respectively), the focal spot of the
lens will be situated around $f \approx 100$~m behind the lens.
This implies that the detector which collects the concentrated 
gamma-rays will be situated on a second spacecraft.
Both spacecrafts fly in formation, but the precision constrains on the 
formation are not very severe.
Due to the small Bragg angles, the focal distance has only to be 
kept within $\pm10$~cm in order to maintain the optimum performances 
of the instrument.
The size of the focal spot is primarily determined by the size of the 
crystal tiles 
(situated between $1\times1$~cm$^2$ and $2\times2$~cm$^2$) and 
the mosaic spread $\Delta \theta$ of the crystals (1 arcmin at a 
distance of 100~m corresponds to a size of 3~cm).
Thus the maximum allowed lateral displacement of the detector 
spacecraft with respect to the lens optical axis will be of the order 
of $\pm1$~cm.
Considering the pointing precision, an accuracy of $\sim15$~arcsec 
are sufficient to maintain the system aligned on the source.

Although the lens is basically a radiation concentrator (with a beam
size that corresponds to the crystal mosaicity, say $\sim1$ arcmin),
it has a substantial off-axis response.
For sources situated off the optical axis, the focal spot will turn 
into a ring-like structure (which is centred on the lens optical 
axis), with an azimuthal modulation that reflects the azimuthal 
angle of the incident photons.
Thus, the arrival direction of off-axis photons can be reconstructed 
from the distribution of the recorded events on the detector plane.
The field-of-view of the lens is therefore basically restricted by 
the size of the detector.
For a detector size of $30 \times 30$~cm$^2$ and a focal length of 
100~m the field-of-view amounts to $\sim15$~arcmin.
Within this field-of-view the lens can be used as (indirect) imaging 
device.
The imaging performances are considerably improved by employing a 
dithering technique, similar to that employed for INTEGRAL.

It is important to notice that the small angle crystal diffraction that 
is exploited for gamma-ray focusing does not alter the polarization of 
the incident radiation.
In other words, a polarized gamma-ray beam will still be polarized 
after concentration on the focal spot, and the use of a polarization 
sensitive detector will allow for polarization measurements
(Caroli et al., these proceedings).
In view of the expected polarization of non-thermal emission, this 
aspect of GRI opens a new discovery space which will considerably 
improve our understanding of the observed objects.

To take maximum profit of the gamma-ray lens, we employ a position 
sensitive detector in the focal spot.
Our actual design studies are mainly focused on a pixelised stack of
detector layers, which on the one hand has the required position 
sensitivity, and on the other hand can be exploited as Compton 
telescope for instrumental background reduction.
Possible detector materials under investigation are CdTe, CZT, Si, 
and/or Ge (Caroli et al. and Wunderer et al., these proceedings).
Although Germanium would provide the best energy resolution (and is
certainly the preferred option for detailed studies of gamma-ray 
lines), the related cooling and annealing requirements may drives us 
towards other options.

The detector configuration we propose (a pixelised detector stack) 
has the nice side effect that it can be employed as an all-sky monitor 
for soft gamma-ray emission (Wunderer et al., these 
proceedings).
If used as a Compton telescope, the detector will be sensitive to any
direction of the sky that is not shielded by the satellite, providing 
thus an onboard capability to detect variable or transient sources.
Since we cannot predict if an all-sky monitor will be available by 
the time that GRI will operate, we think that an all-sky monitoring 
capability aboard GRI itself is important to quickly react to targets 
of opportunity.
We currently investigate technical details of such a solution.

In order to extend the GRI energy coverage towards lower energies
(below $\approx 150$ keV) we also investigate the possibility to add a 
hard X-ray monitor to the mission.
Such a broad-band coverage would be crucial to better understand the 
physics of compact objects, which exhibit spectral variations over
a wide energy band.
In particular, the accurate determination of energy cut-offs will rely 
on an accurate determination of the broad-band spectrum of the object 
under investigation.
Two options are currently considered:
a small multilayer-coated mirror telescope or a collimated coded mask 
telescope (Natalucci, these proceedings).

Theoretically, a multilayer-coated mirror promises an excellent 
sensitivity for hard X-ray astronomy, yet it is unclear up to which 
energies these capacities may be extended.
The american NuStar mission or the French-Italian Simbol-X mission 
plan to use mirrors up to energies of $\sim80$~keV.
Potentially, even higher energies may be reached (e.g.~Christensen,
these proceedings).
Alternatively, a collimated coded mask telescope may provide a good 
option for a hard X-ray monitor on GRI.
At energies $\la100$~keV the cosmic background radiation presents the 
most important source of photons, limiting severely the sensitivity 
of current wide field-of-view coded mask telescopes, such as IBIS on 
INTEGRAL or the BAT on SWIFT).
Collimation considerably reduced this background component, promising
an interesting sensitivity increase at these low energies.

We currently investigate the accommodation of a multilayer-coated 
mirror or a coded mask telescope on GRI.
The mirror could be inserted within the spare area 
($\sim1$~m diameter) inwards of the innermost crystal rings or in 
some specific area reserved on the lens area.
Analogously, the coded mask could have a ring-like shape, and could
be fit within the innermost crystal rings (Natalucci, these 
proceedings).

\section{CONCLUSIONS}

The gamma-ray band presents a unique astronomical window that allows the 
study of the most energetic and most violent phenomena in our Universe.
With ESA's INTEGRAL observatory, an unprecedented global survey of 
the soft gamma-ray sky is currently performed, revealing hundreds
of sources of different kinds, new classes of objects, extraordinary views 
of antimatter annihilation in our Galaxy, and fingerprints of recent 
nucleosynthesis processes.
While INTEGRAL provides the longly awaited global overview over the soft 
gamma-ray sky, there is a growing need to perform deeper, more 
focused investigations of gamma-ray sources, comparable to the 
step that has been taken in X-rays by going from the EINSTEIN  
satellite to the more focused XMM-Newton observatory.
Technological advances in the past years in the domain of gamma-ray 
focusing using Laue diffraction techniques have paved the way towards 
a future gamma-ray mission, that will outreach past missions 
by large factors in sensitivity and angular resolution.
Such a future {\em Gamma-Ray Imager} will allow to study particle 
acceleration processes and explosion physics in unprecedented depth, 
providing essential clues on the intimate nature of the most violent 
and most energetic processes in the Universe.


\end{document}